%% file: main.tex
\documentclass[aps,prl,twocolumn,groupedaddress]{revtex4-2}
\usepackage{blkarray}
\usepackage{graphicx}
\usepackage{amsmath}
\usepackage{hyperref}
\usepackage{physics}
\usepackage{comment}

\newcommand{\zpl}{\mathrm{ZPL}}
\newcommand{\nr}{\mathrm{NR}}

\begin{document}

\title{Overcoming temperature limits in the optical cooling of solids \\ using light-dressed states}

\author{Lu\' \i sa Toledo Tude, Conor N. Murphy and Paul R. Eastham}

\affiliation{School of Physics, Trinity College Dublin, Dublin 2, Ireland}
\affiliation{Trinity Quantum Alliance, Unit 16, Trinity Technology and Enterprise Centre, Pearse Street, Dublin 2, Ireland}

\date{\today}

\begin{abstract}
Laser cooling of solids currently has a temperature floor of 50 -- 100\,K. We propose a method that could overcome this using defects, such as diamond color centers, with narrow electronic manifolds and bright optical transitions.  It exploits the dressed states formed in strong fields which extend the set of phonon transitions and have tunable energies. This allows an enhancement of the cooling power and diminishes the effect of inhomogeneous broadening. We demonstrate these effects theoretically for the silicon-vacancy and the germanium-vacancy, and discuss the role of background absorption, phonon-assisted emission, and non-radiative decay.
\end{abstract}


\maketitle



The optical cooling of solids to low temperatures is an important open challenge. The current method~\cite{epstein_observation_1995, seletskiy_laser_2010,melgaard_solid-state_2016} is anti-Stokes fluorescence in rare-earth doped glasses~\cite{seletskiy_laser_2016}. In this process a rare-earth ion absorbs light, creating an excited electronic state which then absorbs phonons before re-emitting the light at a higher frequency. Despite the competing heating by non-radiative decay and background absorption, temperatures as low as 91\,K have been reached~\cite{melgaard_solid-state_2016}. This is, however, reaching the fundamental limit of 50 -- 100\,K~\cite{seletskiy_laser_2016} set by the phonon energy that can be absorbed effectively. The same issue of a characteristic phonon energy also limits the possible cooling by absorption of optical phonons reported in room-temperature semiconductors~\cite{zhang_laser_2013,zhang_resolved-sideband_2016} (see also Ref.~\cite{morozov_can_2019}), and while various techniques have been considered to improve performance along with different cooling schemes~\cite{zhang_resolved-sideband_2016,seletskiy_laser_2010,melgaard_solid-state_2016, vermeulen_mitigating_2007,rand_raman_2013,andre_saturation_2022}, they do not address this problem. One route to reaching lower temperatures exploits the continuous electronic dispersion in semiconductors~\cite{sheik-bahae_can_2004,rivlin_laser_1997,finkeisen_cooling_1999,rupper_theory_2006,rupper_large_2006}, but has yet to be achieved~\cite{morozov_can_2019}.

\begin{figure}
    \centering
    \includegraphics{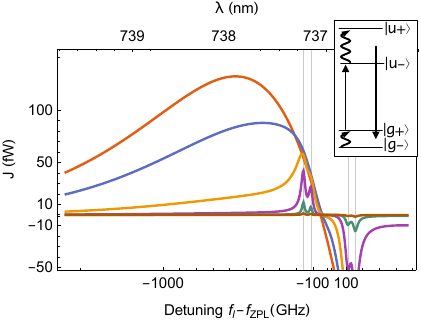}
    \caption{Calculated cooling power of a single laser-driven SiV at $T=20$ K, as a function of laser detuning. Different curves correspond to the different Rabi frequencies, i.e., driving strengths, $\Omega=2\times10^{-0.5},2\times10^{-1},\ldots,2\times10^{-3}\;\mathrm{rad\;ps^{-1}}$. Rabi frequency decreases from the top to the bottom curves on the left of the plot. All polarizations are driven equally, $\Omega_z=\Omega_+=\Omega_-=\Omega$. The vertical lines mark the four optical transitions between the ground-state and excited-state manifolds. The inset shows the spin-up levels of the SiV, and the anti-Stokes cooling process which occurs for weak driving on the lowest-energy transition. The cycle involves phonon transitions (wavy arrows), absorption of photons from the driving laser (straight upward arrow), and spontaneous emission (straight downward arrow). }
    \label{fig:coolingspectra}
\end{figure}

In this Letter, we propose a mechanism by which the temperature floor
of solid-state laser cooling could be overcome, using quasi-resonant
excitation of a suitable defect state. We focus on the group IV color
centers in diamond and in particular the negatively charged
Silicon-Vacancy defect (SiV). The states of this defect comprise a
ground-state and an excited-state manifold with dipole-active optical
transitions between the two\
\cite{hepp_electronic_2014,neu_low-temperature_2013,becker_coherence_2017,bradac_quantum_2019}. At
sufficiently low temperatures the optical spectrum shows four lines,
two of which could be resonantly driven to produce a form of
anti-Stokes cooling. For weak driving this process is very sensitive
to detuning. However, for stronger driving the dynamics becomes
controlled by laser-dressed states rather than the original electronic
eigenstates. The formation of these states, via the
Autler-Townes/a.c. Stark effect~\cite{autler_stark_1955}, leads to a
more complex cooling process with many phonon transitions occurring
within the dressed-state spectrum~\cite{supplement}. This process, which we dub Dressed
state Anti-Stokes Cooling (DASC), corresponds to the inelastic
scattering of the driving laser in the strong-driving
regime. Importantly, it can be tuned to optimize the phonon absorption
rates: by controlling the field intensity and detuning, the energy
gaps and phonon matrix elements are modified in such a way that the
coupling to the phonons is larger for the most occupied states at a
given temperature. For the SiV we predict high gross cooling powers,
of 1 -- 100\,fW per defect, down to temperatures of a few Kelvin. This is
many orders-of-magnitude greater than could be achieved using
rare-earth ions, even at the higher temperatures where they can
operate. Furthermore, the cooling effect occurs over a broad range of
detunings (see Fig. \ref{fig:coolingspectra}), and so survives in the
presence of inhomogeneous broadening. While it will compete with various heating processes, as we discuss later, our results suggest DASC could enable the optical cooling of solids to temperatures unachievable with existing methods.

The SiV in diamond~\cite{becker_coherence_2017} is one of a family of group IV color centers~\cite{bradac_quantum_2019} that are candidates for DASC. We focus on the SiV, although we obtain similar results for the Germanium-Vacancy (GeV)~\cite{supplement}. These defects are of interest for quantum nanophotonics, acting as both optically addressable spin qubits and effective single-photon emitters~\cite{neu_single_2011}. The electronic and spin states have long coherence times in the low temperature regime we consider, producing lifetime-limited spectra without spectral diffusion~\cite{becker_coherence_2017}. The spin coherence time for the SiV is sub-microsecond due to phonon scattering within the ground-state manifold~\cite{becker_coherence_2017}. While this is disadvantageous for use as a spin qubit, it is beneficial for cooling because it allows rapid phonon absorption. Conventional anti-Stokes cooling using the SiV and related Nitrogen Vacancy (NV) was considered in recent works~\cite{kern_optical_2017,gao_phonon-assisted_2018}, based on the broad absorption lines observed at room temperature. Here we consider, instead, high quality materials at low temperatures. This opens up the possibility of precision cooling approaches with tailored excitation of well-characterized transitions. Methods for cooling micromechanical resonators using the NV and SiV have been proposed~\cite{kepesidis_phonon_2013,kepesidis_cooling_2016}, however, the goal in that case is to cool a single resonator mode, whereas we consider cooling of the bulk material.

A disadvantage of our method is that it will require pre-cooling of the sample, unlike anti-Stokes cooling which operates from room temperature. Our analysis is restricted to a low temperature regime where we may use the Born-Markov approximation including only lowest-order single-phonon processes~\cite{supplement}, and where heating due to non-radiative decay may not be overwhelming. For the SiV there is a strong decrease in the emission lifetime~\cite{jahnke_electronphonon_2015} above $\sim 100\,\mathrm{K}$, suggesting significant non-radiative decay which would preclude cooling. Another disadvantage of this type of defect is the presence of emission into phonon sidebands~\cite{becker_coherence_2017,neu_single_2011, londero_vibrational_2018, gorlitz_spectroscopic_2020}. This will produce a competing heating effect, which would need to be suppressed using photonic structures~\cite{ondic_photonic_2020,riedrich-moller_deterministic_2014,ruf_resonant_2021}.




To study optical cooling using an SiV we use an open-quantum systems description based on the Born-Markov approximation.  As usual, we divide the problem into a system and baths, with the system Hamiltonian describing the SiV driven by a laser. It is a one-hole system with eight levels, four associated with spin-up and four with spin-down. The levels in each spin sector form an excited (u) and ground (g) state manifold, each containing two levels (see Fig.~\ref{fig:coolingspectra}). The states in each manifold, $|u_{\pm}\rangle$ and $|g_{\pm}\rangle$, are split by the spin-orbit coupling constants $\lambda^u=1.11$ meV and $\lambda^g=0.19$ meV  \cite{hepp_electronic_2014}. We ignore the small contribution of the linear vibronic Jahn-Teller interaction, which is 
around 5\% of the spin-orbit coupling and so would not qualitatively affect our results. The manifolds are separated by the Zero Phonon Line (ZPL) energy $E_{\zpl} =1.68$ eV, and coupled by electric-dipole optical transitions of various polarizations. We suppose that these transitions are driven by lasers of a single frequency $\omega_l$, and use the Floquet representation obtained after a unitary transformation $U=e^{i\omega_l t (|u_+\rangle+|u_-\rangle)}$. The Hamiltonian for the spin-up sector is, with $\hbar=1$, 
\begin{equation}
H_{S}=
 \frac{1}{2}\begin{pmatrix}
    -\Delta+\lambda^u & 0 & \Omega_z & \Omega_+ \\
    0 & -\Delta-\lambda^u & \Omega_- & \Omega_z \\
    \Omega_z & \Omega_- & \Delta+\lambda^g & 0 \\
    \Omega_+ & \Omega_z & 0 & \Delta-\lambda^g\\ 
  \end{pmatrix}.
\end{equation} The basis here is in the order of decreasing energy in the original frame. In the Floquet basis the energies are shifted by the laser frequency, and given in terms of the detuning from the zero-phonon line $\Delta=\omega_l-\omega_{\zpl}$. $\Omega_{z}$ is the Rabi frequency quantifying the strength of the driving with polarization along the z-axis, which is the [111] crystal axis, and $\Omega_{\pm}$ are the Rabi frequencies for the two circular polarizations in the xy plane. The Hamiltonian for the spin-down sector is identical, except that the two circular polarizations are swapped. Most of our results refer to situations where $\Omega_{+}=\Omega_{-}$, so that $H_S$ describes both spin sectors.

Acoustic phonons cause transitions between states in each manifold due to the intramanifold coupling~\cite{kepesidis_cooling_2016} 
\begin{equation}
    H_{I1}=\sum_{k} \left(f^u_k\ket{u_+}\bra{u_-} + f^g_k\ket{g_+}\bra{g_-}  +\mbox{H.c}.\right)\otimes (a_k + a^{\dagger}_k).
\end{equation} They also affect the energy separation of the manifolds due to the intermanifold deformation-potential coupling~\cite{norambuena_quantifying_2020,norambuena_microscopic_2016} 
\begin{align}
\begin{split}
H_{I2}=\sum_{k} g_k 
&\left(\ket{u_+}\bra{u_+}+\ket{u_-}\bra{u_-}\right.\\
&\left.-\ket{g_+}\bra{g_+}-\ket{g_-}\bra{g_-}\right)\otimes\left(a_k+a^{\dagger}_k\right).
\end{split}
\end{align} The bath has a super-Ohmic spectral density, reflecting the density-of-states of the acoustic phonons and the frequency dependence of the electron-phonon coupling. We use the expression \cite{norambuena_microscopic_2016,norambuena_quantifying_2020} $J_p(\omega)=2A\omega_c^{-2}\omega^3 e^{-\omega/\omega_c}$ with $A=0.0275$ and $\omega_c=2\pi\;\mathrm{rad\;ps^{-1}}$. We include only the bulk phonon contribution relevant to our results, neglecting the local phonons which have much higher energies. The intermanifold coupling has been modelled~\cite{kepesidis_cooling_2016} using the same form of $J_p(\omega)$ but with $A=0.073$, reflecting the difference in matrix elements. We include this difference in the ratio $f^u_k/g_k$ and for simplicity take $f_k^u=f_k^g$.

An equation-of-motion for the reduced density matrix of the SiV can be obtained using the Born-Markov approximation~\cite{breuer_theory_2007}, which is well justified in the parameter regimes we consider~\cite{supplement}. The general form we use here is given in~\cite{murphy_laser_2022,supplement}, along with the form for the mean heat current, from the phonons to the SiV, obtained using the method of full counting statistics~\cite{esposito_nonequilibrium_2009}.  The contribution to the heat current from phonon absorption is $J_{ab} = \sum_{ij} J_{ij}$,
\begin{equation}J_{ij} \propto \nu_{ji} n(\nu_{ji})J_p(\nu_{ji}),\label{eq:J}\end{equation} where $\nu_{ji}=E_j-E_i$ are transition energies between the eigenstates of $H_S$,  $J_p(\omega)$ is the phonon spectral density, and $n(\omega)=(e^{\beta\omega}-1)^{-1}$ the occupation. We include the radiative decays using standard Lindblad dissipators~\cite{breuer_theory_2007}, one for each transition from one of the $u$ states to one of the $g$ states. We neglect differences in the rates of the different decays, which are factors of order one, and take them all to have rate $\gamma_0=1\;\mathrm{ns^{-1}}$\cite{sternschulte_1681-ev_1994,wang_single_2005}. We drop the Lamb shift terms, which are negligible, but do not make the secular approximation. 

Fig.~\ref{fig:coolingspectra} shows the calculated cooling power as a function of driving frequency, for a single SiV at a temperature of 20\,K, driven with $\Omega_z=\Omega_+=\Omega_-$. Different curves correspond to different driving strengths, i.e., Rabi splittings, similar to those reached in experiments~\cite{zhou_coherent_2017}. At weak driving, we see two characteristic peaks on each side of zero detuning, with those on the red (blue) side corresponding to net cooling (heating) of the phonons. The cooling (heating) effect corresponds to weak resonant driving of the two lowest (highest) of the four lines in the optical spectrum. As the driving is increased the red-detuned features first evolve into a broad spectral region where there is a very high cooling power. Further increases, beyond the driving strengths shown, then lead to a reduction in the maximum power. 

\begin{figure}
    \centering
    \includegraphics{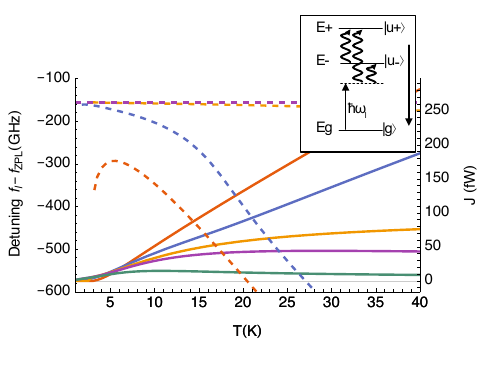}
    \caption{Maximum cooling power (solid curves, right axis) and corresponding detuning (dashed curves, left axis) as a function of temperature for several driving strengths. Results are shown for $\Omega_z=\Omega_+=\Omega_-=2\times10^{-0.5},2\times10^{-1},\ldots,2\times10^{-2.5}\;\mathrm{rad\;ps^{-1}}$. The smallest Rabi frequency corresponds to the smallest power on the right of the plot, and the largest (least negative) detuning. The detuning curves for the two smallest Rabi frequencies coincide. The inset illustrates the quasienergies for a three-level model and the DASC cycle.}
    \label{fig:tempandtunings}
\end{figure}

Figure~\ref{fig:tempandtunings} shows the calculated maximum cooling power, obtained by maximizing over the detuning, as a function of temperature. We also show the corresponding values of the detuning. For weak driving, the optimal detuning is indistinguishable from that of the lowest-energy transition $|g_+\rangle\rightarrow |u_-\rangle$, $-(\lambda^u+\lambda^g)/2$, and is independent of temperature. This is consistent with the anti-Stokes cooling process illustrated in the inset to Fig.\ \ref{fig:coolingspectra}. For the lowest temperatures shown this process involves phonon transitions only in the ground-state manifold, giving a small cooling power. The cooling power increases as temperature is increased, allowing phonon absorption in the upper manifold, but then becomes constant, due to the presence of a characteristic maximum energy, $\lambda^u$, that can be absorbed per cycle at weak driving. This saturation is absent in the strong-driving case, where the cooling power increases linearly with temperature. The optimal detuning is lower (more negative) than in the weak-driving case, corresponding to driving on the red side of the lowest-energy transition, and is strongly temperature dependent, becoming more negative as temperature increases.

The enhancement of the cooling power arises because for strong driving the heat flow to the phonons is controlled by transitions between the laser-dressed states, whose quasienergies are the eigenvalues of $H_S$. As can be seen in Eq. (\ref{eq:J}), each pair of states $i,j$ gives rise to a heat flow proportional to the transition frequency, $\nu_{ji}$, and the Bose function and spectral density at that frequency. Thus, by tuning the quasienergies through the driving field, we can increase the heat absorption beyond the maximum achievable with a fixed level structure. 

In the four-level system there are multiple phonon absorption pathways, leading to a complex cooling cycle discussed further in the Supplemental Material~\cite{supplement}. However, we obtain similar results for a simplified three-level model with a single ground state and two excited states. This allows us to identify the essential features of the DASC process as those illustrated in the inset of Fig.\ \ref{fig:tempandtunings}. The eigenstates of the defect, in the presence of the periodic driving, are the Floquet states, with quasi-energies defined up to integer multiples of the driving frequency. This introduces a periodic replica of the ground-state, near to the excited states. The driving field then mixes the states together. Referring to Fig.\ \ref{fig:tempandtunings} we see that, over the temperature regime shown, the driving strength is a fraction of the detuning, so the mixing produces only small energy shifts. However, it gives rise to the two additional cooling processes, in which a photon and phonon are simultaneously absorbed. In terms of the original (undressed) states, these are Raman transitions from the ground to the excited states via a virtual intermediate state. The energy of the absorbed phonon in the two cases is approximately $(E_{\mp}-E_{g})-\hbar \omega_l$, and can be tuned to maximize the cooling power. The increase in cooling power in Fig.\ \ref{fig:tempandtunings} over the weak-driving case in the regime $kT\gtrsim \lambda^u$ arises because the larger transition energies increase the heat absorption. The transition energies can also be smaller than those in the bare spectrum, allowing cooling at temperatures below the limit set by the bare level splitting. Indeed, as can be inferred from the illustration, DASC can occur in a two-level system.

There is a maximum in the cooling power as a function of driving strength because the relaxation processes, and therefore the heat flows, depend on the dressed-state energies and compositions~\cite{geva_three-level_1994}. At large $\Omega$ the phonon heat current changes sign, and the cycle operates as a heater. For a two-level model one can define an effective temperature for the photons, whose relation to temperature of the phonon bath determines the direction of heat flow~\cite{murphy_quantum_2022}. 

The cooling will compete with heating due to background absorption in the host material. The Rabi frequency $\hbar \Omega=d E_0$, where $d$ is the transition dipole moment and $E_0$ the electric field amplitude, and the intensity is $I=c\epsilon_0nE_0^2/2$. The heating rate due to background absorption, per SiV, is then $c\epsilon_0n\hbar^2\Omega^2\alpha_b/(2d^2\rho)$, where $\alpha_b$ is the background absorption coefficient and $\rho$ the SiV density. Considering the strongest field used in Fig.~\ref{fig:coolingspectra}, we have a heating power $\sim 10^{25}(\alpha_b/\rho)\;\mathrm{fW\; m^{-2}}$, against a cooling power of $\sim 100\;\mathrm{fW}$. Taking a conservative background absorption coefficient $\alpha_b=0.1\;\mathrm{cm^{-1}}$~\cite{thomas_multiphonon_1994}, and $d=14.3$ Debye~\cite{becker_coherence_2017} we then predict net cooling for $\rho\gtrsim 10^{24}\;\mathrm{m^{-3}}$, or about one SiV per $10^5$ carbon atoms. Thus background heating is unlikely to be a limitation.

A second competing heating effect arises from radiative decay accompanied by the emission of phonons. In general, only a fraction $\alpha$ of the radiative decay is into the ZPL, with the remainder into the phonon sideband. For DASC, the maximum cooling power is limited by the radiative decay rate into the ZPL, and so is $\sim kT \alpha \gamma_0 p_u$, where $p_u$ is the probability of occupying the upper manifold. The heating rate by sideband emission, with mean phonon energy $\bar{E}_{sb}$, is $\sim (1-\alpha) \gamma_0 p_u \bar{E}_{sb}$, so net cooling requires $\alpha>\bar{E}_{sb}/(\bar{E}_{sb}+kT)$. For the SiV, there is a strong effect of a local phonon with energy $\approx 60\,\mathrm{meV}$~\cite{norambuena_microscopic_2016}, so at $30\,\mathrm{K}$ net cooling occurs if $\alpha>0.95$. Many works~\cite{londero_vibrational_2018,rose_observation_2018,bradac_quantum_2019,jahnke_electronphonon_2015,hausler_photoluminescence_2017} suggest $\alpha \approx 0.7$, however, $\alpha\approx 0.9$ has been reported~\cite{becker_coherence_2017,neu_single_2011}. Furthermore, $\alpha$ can be increased by using photonic structures~\cite{ondic_photonic_2020,riedrich-moller_deterministic_2014,ruf_resonant_2021}.

A third competing heating effect is non-radiative decay from the $u$ to the $g$ manifold. The impact of this will be similar to that in conventional laser cooling~\cite{seletskiy_laser_2016}. A non-radiative decay with rate $\gamma_{\nr}$ implies heating $E_{\zpl} \gamma_{\nr} p_u$. Net cooling at $10$ K then requires $(\gamma_0 + \gamma_{\nr})/\gamma_{\nr}>E_{\zpl}/kT\gtrsim 1000$, or a quantum efficiency (QE) of at least $99.9\%$. Experimental results for the QE in the low-temperature regime relevant here are scarce, although a lower bound on the QE of $30\%$ has been found~\cite{becker_coherence_2017} from the measured lifetime and dipole moment. However, significantly longer lifetimes have been measured~\cite{sternschulte_1681-ev_1994}, which would imply much higher QEs, as would the rates from fits to second-order coherence data~\cite{neu_low-temperature_2013}. A value of $67\%$ was found in room-temperature experiments~\cite{riedrich-moller_deterministic_2014}, again implying a high QE at low temperatures, where the activated non-radiative processes disappear. Given these results we suggest that the QE required for cooling is achievable, with the low values in the literature perhaps due to extrinsic effects such as strain. 
We note also recent measurements of the Germanium vacancy~\cite{bhaskar_quantum_2017,palyanov_germanium_2015,iwasaki_germanium-vacancy_2015} which suggest a high QE. We have investigated cooling in this defect and obtain results similar to those for the SiV~\cite{supplement}.

\begin{figure}
  \centering
    \includegraphics{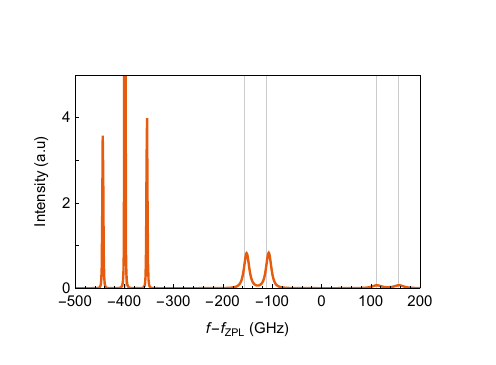}
    \caption{Computed spectrum of the light from an SiV at 20\,K driven with $\Omega_+=\Omega_-=\Omega_z=2\times 10^{-1}\,\mathrm{rad\;ps^{-1}}$. This corresponds to the second-highest driving shown in Fig.~\ref{fig:coolingspectra}. The laser is detuned to the maximum in the cooling power at $f_l-f_{\zpl}=-400$ GHz. All polarization components have been combined to show the total power.}
    \label{fig:emspec}
\end{figure}

The DASC process could be observed via the scattering of the driving laser. Fig.~\ref{fig:emspec} shows the spectrum of the light from an SiV driven in the cooling regime. It contains two sidebands around the laser frequency, forming a Mollow triplet~\cite{mollow_power_1969}, and four broader peaks near to the original emission lines. The scattered intensity at frequencies above the driving, at $f_l=-400$ GHz, is greater than that below it, so on average the scattering increases the energy of the photons. The energy they gain corresponds, by energy conservation, to that lost by the phonons, and this blueshifting of the scattering demonstrates that heat is being removed from the crystal. 

In conclusion, we have proposed and analyzed an approach to the optical cooling of solids using quasi-resonant excitation of optically-active defect states in a low-temperature regime. Our approach could overcome the temperature limitations inherent in optical cooling using rare-earths, set by the gaps within the electronic manifolds, by using laser dressing to control the frequencies of the phonon transitions. This allows the process to take advantage of the maximum in the phonon heat absorption rate for transition energies $\sim kT$ and so produce a large cooling power. This could allow systems such as the group-IV color centers in diamond to cool effectively. The dressed-state enhancement, along with the fast radiative rates, leads to predicted cooling powers many orders of magnitude greater than could be obtained using rare-earths, down to temperatures of a few Kelvin. An interesting question is the maximum temperature at which these processes could be effective. While the Born-Markov theory will eventually break down, it is likely that the upper limit in practice will be set by the breakdown of the four-level model and the absorption and decay arising from other states~\cite{supplement}. 


\begin{acknowledgments}
We acknowledge funding from the Irish Research Council under award GOIPG/2017/1091, and Science Foundation Ireland (21/FFP-P/10142). 
\end{acknowledgments}

L.T.T. and C.N.M. contributed equally to this work.

\nocite{popovic_quantum_2021,gauger_heat_2010,murphy_quantum_2019}
%

\end{document}



\title{Supplemental Material for ``Overcoming temperature limits in the optical cooling of solids using light-dressed states''}

\author{Lu\' \i sa Toledo Tude, Conor N. Murphy and Paul R. Eastham}

\affiliation{School of Physics, Trinity College Dublin, Dublin 2, Ireland}

\date{\today}

\maketitle


This supplemental material is organized as follows. In Sec.\ \ref{sec:suppmethods} we provide a summary of the computational methods. These methods are implemented in the accompanying Mathematica notebook which was used to generate the results. In Sec.\ \ref{sec:suppsiv} we give some additional discussion of the cooling process. In Sec.\ \ref{sec:suppbm} we quantify the validity of the Born-Markov approximation. In Sec.\ \ref{sec:suppgev} we present results for the Germanium vacancy in diamond and compare them to those for the SiV given in the main text. 

\section{Methods}
\label{sec:suppmethods}

The results, here and in the main text, are obtained using the Bloch-Redfield equation and corresponding expressions for heat currents from Ref.~\cite{murphy_laser_2022}, which we recall briefly here. As usual in the treatment of an open quantum system, we separate the full Hamiltonian into a system, $H_S$, a bath of harmonic oscillators, $H_B$, and a system-bath interaction, $H_{SB}=\sum_k g_k O (a_k+a_k^\dagger)$. $O$ is the system operator to which the bath couples, $a_k$ the annihilation operator of the $k$-th mode of the bosonic bath, and $g_k$ the coupling strength. Using the Born and Markov approximations we can obtain an equation-of-motion for the reduced density matrix of the system, $\rho_S(t)=\Tr_{B} \rho(t)$, \begin{equation}\frac{d\rho_S(t)}{dt}=-i[H_S,\rho_S(t)]+
\sum_{ij}\big\{ A_{ij} [O_{ji}\rho_S(t)O+O\rho_S(t)O_{ij} 
-\rho_S(t)O_{ij}O-O O_{ji}\rho_S(t) ] \big\}.\label{eq:suppnsmaster}\end{equation}
The operators $O_{ij}$ are defined by the representation of $O$ in the eigenbasis of $H_S$, \begin{equation}
  O=\sum_{ij} \langle i|O|j\rangle |i\rangle\langle
  j| \equiv \sum_{ij} O_{ij}.\end{equation} The principal value terms have been neglected. The quantities $A_{ij}$ are the transition rates \begin{equation} A_{ij}=\pi \{[n(\nu_{ij})+1]J_p(\nu_{ij})+n(\nu_{ji})J_p(\nu_{ji})\}.\label{eq:aij}\end{equation} $\nu_{ij}=E_i-E_j$ are the transition frequencies, $n(\nu)$ the Bose function, and $J_p(\nu)$ is the spectral density  of the bath, which is zero for $\nu<0$. We can also obtain, using these approximations and the method of counting statistics, the average heat current between the system and the bath, \begin{equation} \frac{d\langle Q\rangle}{dt}=\sum_{ij}\big\{ A_{ij}  [(E_i-E_j)\Tr (O_{ji}\rho_{S}(t)O+O\rho_{S}(t)O_{ij}] \big\}.\label{eq:suppheatcurrent}\end{equation}

The results are obtained using Eq. \ref{eq:suppnsmaster} to describe the dynamics and the interaction with phonons, with additional Lindblad dissipators for spontaneous emission. We use the Floquet representation throughout with the system Hamiltonian given by Eq. 1 in the main text. The eigenvectors and eigenvalues of $H_S$ are computed numerically, and the steady-state density-matrix determined from the master equation. This is then used to compute the heat currents according to Eq.~\ref{eq:suppheatcurrent}. The Mathematica notebook used to perform these calculations is provided in the file {\tt{coolingcalculations.nb}} which forms part of this supplemental material.

The principles behind the advantage of our method can be understood from the analytical expressions presented in this section. Generally, at low temperatures, the occupation number $n(\nu_{ij})$ assumes small values, preventing the cooling process. DASC overcomes this issue since under strong driving the energy gaps $\nu_{ij}$ are altered, and can be tuned (by changing the detuning and driving strength) to optimize the product $\nu n J_p$ that controls the heat absorption rates via Eqs.~\ref{eq:aij} and \ref{eq:suppheatcurrent}. In addition, strong driving expands the number of transitions accounted in the sum of Eq.~\ref{eq:suppheatcurrent}, i.e, it provides additional cooling channels.




\section{Additional results for the SiV}

\label{sec:suppsiv}

\begin{figure}
    \centering
\includegraphics{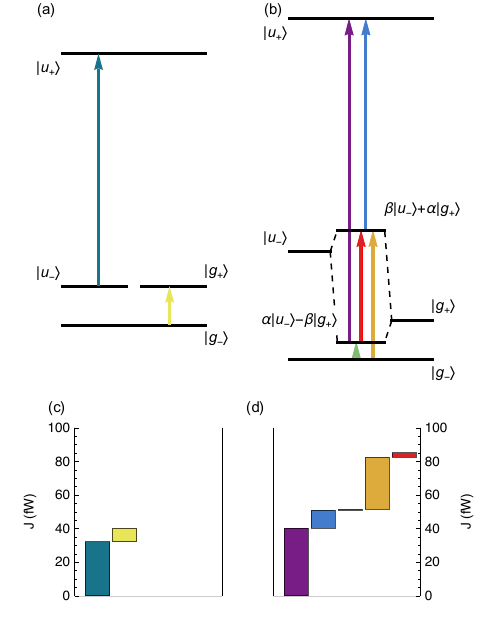}
\caption{Energy levels of the Floquet states (top row), and associated heat currents for each pair of levels (bottom row) for weak resonant driving (left column), and strong driving (right column). For weak driving the energy levels are unaffected, and phonon cooling occurs via transitions between the unperturbed states. Strong driving gives the energies shown in the central part of panel (b), which are shifted relative to the undriven case (outer parts). Phonon cooling occurs via transitions between laser-dressed states formed from admixtures of the original levels. $\Omega_+=\Omega_z=0$. Left column: $\Omega_-=2\times10^{-2}\;\mathrm{rad\;ps^{-1}}$, with driving on the $|g_+\rangle\rightarrow|u_-\rangle$ transition. Right column: $\Omega_-=2\times10^{-0.5}\; \mathrm{rad\;ps^{-1}}$, with driving $80$ GHz below the transition.}
\label{fig:levelsandbridgeplot}
\end{figure}

The cooling processes in the four-level system are somewhat complex due to the presence of multiple transitions. To elucidate them we show, in Fig.~\ref{fig:levelsandbridgeplot}, the energy levels for $H_S$, and the breakdown of the heat currents into contributions from each pair of levels $J_{ij}$. Each of these contributions arises from both emission and absorption processes, related to the first and second terms in Eq. \ref{eq:aij}, respectively. For simplicity, we have taken only $\Omega_-$ to be non-zero. Panels (a) and (c) illustrate the behavior for weak driving resonant with the transition from $|g_+\rangle$ to $|u_-\rangle$. In the Floquet frame, these levels are degenerate, and weakly coupled by the driving laser. This does not produce a noticeable energy splitting, and its effect is merely to transfer the population between the states. This leads to the standard anti-Stokes-type cooling cycle operating with the unperturbed states: phonons are absorbed in transitions within each manifold, and the cycle is closed by absorption of laser photons and radiative decay. The energy shifts and line broadenings are small, and cooling only occurs close to resonance. 


The behavior for stronger driving, with detuning below the lowest-energy transition, is exemplified by Fig.~\ref{fig:levelsandbridgeplot}b. This diagram shows the original energy levels in the Floquet frame, $|u_-\rangle$ and $|g_+\rangle$, as well as the energy levels of $H_S$. The original states are mixed by the driving field, to form dressed-states $|D_{\pm}\rangle=\beta|u_-\rangle\pm\alpha|g_+\rangle$. The energy shifts, due to the Autler-Townes effect, can be seen. The mixing allows phonon processes which do not occur otherwise. These include transitions from the state nearest in energy to $|g_+\rangle$, $|D_{-}\rangle$, to the state $|u_+\rangle$, which can occur via $H_{I1}$ given the admixture of $|u_-\rangle$ in $|D_{-}\rangle$, and transitions between the dressed states, $|D_\pm\rangle$, shown by the red arrow. This second process~\cite{gauger_heat_2010,murphy_quantum_2019} was previously predicted to allow cooling for quantum-dot excitons under the deformation-potential interaction, though here it provides only a small fraction of the heat current. As can be seen in Fig.~\ref{fig:levelsandbridgeplot}d, the increase of the heat current for strong driving is mostly due to the presence of the additional transitions in purple and yellow, which reflects the match between these transition energies and the peak in the heat absorption rates. Since significant mixing occurs for driving within a detuning $\sim \Omega$ of a resonance, the cooling effect does not require exact resonance and will survive some inhomogeneous broadening.  Referring to Fig.~1 of the main text we see that, for the largest Rabi splitting shown, cooling would be possible for inhomogeneous broadenings up to several nanometers. This is similar to that seen in polycrystalline diamond, and much larger than that in the monocrystalline material~\cite{neu_low-temperature_2013,sternschulte_1681-ev_1994}.

\section{Validity of Born-Markov approximation}
\label{sec:suppbm}

\begin{figure}
    \centering
    \includegraphics[width=.99\linewidth]{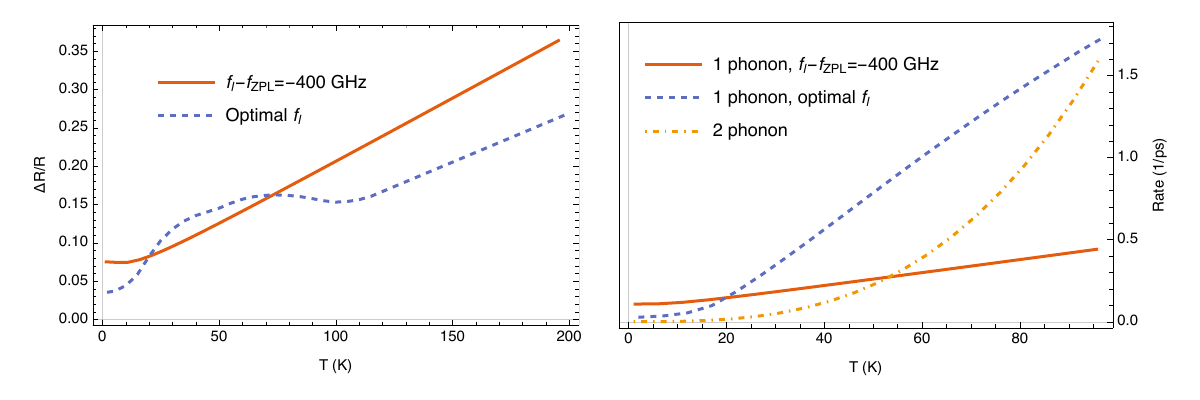}
    \caption{Left panel: Maximum spectral variation of the rates
    $\Delta R/R$ as a function of temperature for the model silicon vacancy, with driving $\Omega_z=\Omega_+=\Omega_-=0.2\;\mathrm{rad\;ps^{-1}}$. The solid orange curve shows the results with fixed detuning $f_l-f_\zpl=-400\,\mathrm{GHz}$. The dashed blue curve shows results with the detuning set to maximize the cooling power at each temperature. Right panel: Comparison of single-phonon transition rates with the two-phonon decoherence rate. The solid orange curve and blue dashed curves are the maximum rates of the single-phonon processes, for the two cases shown in the left panel. The yellow dot-dashed curve is the two-phonon decoherence rate in the excited-state manifold for the undriven vacancy.}
    \label{fig:bornmark}
\end{figure}

The Born-Markov approximation, used to obtain Eqs.\ \ref{eq:suppnsmaster} and \ref{eq:suppheatcurrent}, is valid only when the interaction-picture density matrix $\tilde\rho_S(t)$ is approximately constant over the timescale on which the bath correlation functions decay. In the frequency domain, this means that the incoherent rates $A_{ij}$ should vary little over the frequency range set by these same rates, i.e., \begin{equation}\Delta R/R=\max_{i,j}\left|\frac{1}{A_{ij}}\frac{d A_{ij}}{d\nu_{ij}}\right| \ll 1.\label{eq:bmvalidity}\end{equation} In the left panel of Fig.~\ref{fig:bornmark} we plot the variation $\Delta R/R$ as a function of temperature, for the driven SiV model. We see that the Born-Markov approximation is reasonably well justified at low temperatures, but is expected to become increasingly inaccurate as the temperature increases. 

In addition, the Born-Markov approximation is second-order in the defect-phonon coupling strength, and as such only includes processes corresponding to the emission or absorption of a single phonon. However, there are also two-phonon processes, where a phonon elastically scatters from the defect, that produce decoherence~\cite{jahnke_electronphonon_2015}. The corresponding decoherence rate varies as $T^3$, whereas the single-phonon processes vary as $T$, so the two-phonon process is the dominant cause of line broadening at high temperatures. This effect is shown in the right-hand panel of Fig.\ \ref{fig:bornmark}, where we compare the single-phonon rates from our theory with the two-phonon decoherence rate for the undriven defect, computed using Eq. 9 in Ref.~\cite{jahnke_electronphonon_2015}, with the spectral density given in our main text. From this comparison we expect our results to be reasonably accurate below around $50\,\mathrm{K}$, depending to some extent on the other parameters. In principle the cooling may survive the presence of decoherence, and indeed corrections to the Markovian approximation, and it would be interesting to explore these effects in future work. This could be done using numerically exact path-integral calculations of the heat  currents~\cite{popovic_quantum_2021}. In any case, however, the non-radiative decay which sets in above $\sim 100\mathrm{K}$ in the SiV~\cite{jahnke_electronphonon_2015} will limit the application of DASC to lower temperatures.

\section{Germanium vacancy in diamond}
\label{sec:suppgev}

In this section, we investigate the performance of the proposed cooling method in another color center in diamond. The Germanium-vacancy (GeV) center is a suitable candidate as it has a level structure similar to the SiV but with larger spin-orbit splitting, and a high quantum efficiency~\cite{bhaskar_quantum_2017, palyanov_germanium_2015}. This defect is less well characterized than the SiV, and we were unable to find some parameters of our model in the literature. Hence we consider only the effects of changing the intra-manifold level splittings on the cooling, using the values for GeV $\lambda^u=4.5$\,meV and $\lambda^g=0.7$\,meV, and taking the remaining parameters to be those for the SiV given in the main text.

\begin{figure}
    \centering
    \includegraphics[width=.99\linewidth]{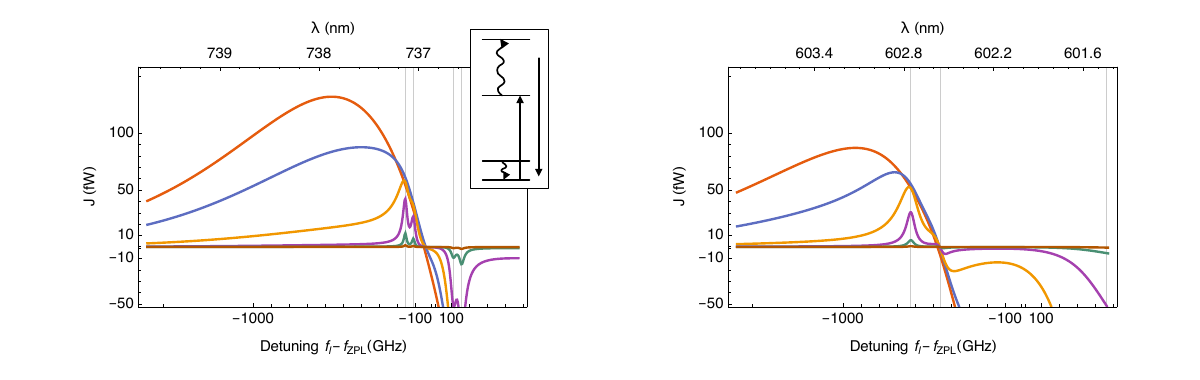}
    \caption{Calculated cooling power of a single laser-driven SiV (left) or GeV (right) at $T=20$ K, as a function of laser detuning. Different curves correspond to the different Rabi frequencies, i.e., driving strengths, $\Omega=2\times10^{-0.5},2\times10^{-1},\ldots,2\times10^{-3}\;\mathrm{rad\;ps^{-1}}$. Rabi frequency decreases from the top to the bottom curves on the left of the plots. All polarizations are driven equally, $\Omega_z=\Omega_+=\Omega_-=\Omega$. The vertical lines mark transitions between the ground-state and excited-state manifolds. The inset in the left panel shows the levels and transitions for weak driving of the SiV on the second-lowest absorption line, as discussed in the supplemental text.}
    \label{fig:1geV}
\end{figure}

In Figure \ref{fig:1geV} we show the cooling power for the GeV as a function of detuning for different Rabi frequencies at $20$\,K, along with the corresponding results from the main text for the SiV. The peak cooling power for the GeV is slightly lower than for the SiV, however, the transition from weak to strong driving is observed similarly. A noticeable difference is that in the weak driving regime there are two cooling resonances for the SiV, but only one for the GeV. The reason for this can be seen in the inset to the left panel, where we illustrate the level structure and the thermodynamic cycle for weak driving of the SiV on the second-lowest absorption line. The levels here are indicated in the original basis before the transformation to the rotating frame. The laser driving is shown as the upwards pointing solid arrow, and the radiative decay as the downwards one. Given this driving, phonons are emitted in the ground-state manifold, but absorbed in the excited-state one, as indicated by the wavy arrows. The heat current due to the latter exceeds that due to the former, so the net effect is to cool the phonons. In the GeV, however, the splitting of the excited-state doublet is much greater than $kT$ at 20\,K, so there is no phonon absorption there. Only the phonon transitions within the ground-state manifold remain, giving net heating. 

\begin{figure}
    \centering
    \includegraphics[width=.99\linewidth]{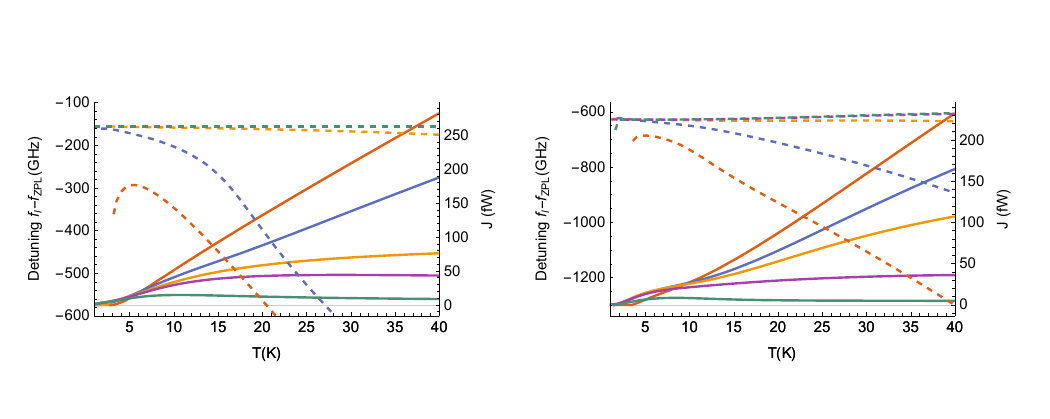}
    \caption{Maximum cooling power (solid curves, right axes) and corresponding detuning (dashed curves, left axes) as a function of temperature for several driving strengths for the SiV (left) and GeV (right). Results are shown for $\Omega_z=\Omega_+=\Omega_-=2\times10^{-0.5},2\times10^{-1},\ldots,2\times10^{-2.5}\;\mathrm{rad\;ps^{-1}}$. The smallest Rabi frequency corresponds to the smallest power on the right of the plots, and the largest (least negative) detuning. The detuning curves for the two smallest Rabi frequencies coincide.}
    \label{fig:3geV}
\end{figure}

Fig. \ref{fig:3geV} shows the maximum cooling power and optimal detuning in the GeV for different temperatures, again with the corresponding results for the SiV from the main text. As discussed in the main text, under weak driving the heat current becomes constant above a certain temperature, reflecting the maximum phonon energy which can be absorbed per cycle.  For the SiV this corresponds to the splitting of the excited-state manifold, $\lambda^u$. For the GeV there is no phonon absorption on that manifold at this temperature, and the saturation temperature corresponds to the splitting of the ground-state manifold, $\lambda^g/k\approx 7$ K. As in the SiV, strong-driving in the GeV allows more energy absorption per cycle above this saturation temperature, producing a much greater cooling power. 

\newsavebox\mytempbib
\savebox\mytempbib{\parbox{\textwidth}{\input{main.bbl}}}